
\font\tenmib=cmmib10
\font\sevenmib=cmmib10 scaled 800
\font\titolo=cmbx12
\font\titolone=cmbx10 scaled\magstep 2
\font\cs=cmcsc10

\font\ninerm=cmr9
\font\ottorm=cmr8
\textfont5=\tenmib
\scriptfont5=\sevenmib
\scriptscriptfont5=\fivei

\font\msytw=msbm9 scaled\magstep1

\font\msytwww=msbm7 scaled\magstep1
\font\indbf=cmbx10 scaled\magstep2

\newskip\ttglue
\font\ottorm=cmr8\font\ottoi=cmmi8\font\ottosy=cmsy7
\font\ottobf=cmbx7\font\ottott=cmtt8\font\ottosl=cmsl8\font\ottoit=cmti7
\font\sixrm=cmr6\font\sixbf=cmbx7\font\sixi=cmmi7\font\sixsy=cmsy7
\font\fiverm=cmr5\font\fivesy=cmsy5\font\fivei=cmmi5\font\fivebf=cmbx5

\def\ottopunti{\def\rm{\fam0\ottorm}\textfont0=\ottorm%
\scriptfont0=\sixrm\scriptscriptfont0=\fiverm\textfont1=\ottoi%
\scriptfont1=\sixi\scriptscriptfont1=\fivei\textfont2=\ottosy%
\scriptfont2=\sixsy\scriptscriptfont2=\fivesy\textfont3=\tenex%
\scriptfont3=\tenex\scriptscriptfont3=\tenex\textfont\itfam=\ottoit%
\def\it{\fam\itfam\ottoit}\textfont\slfam=\ottosl%
\def\sl{\fam\slfam\ottosl}\textfont\ttfam=\ottott%
\def\tt{\fam\ttfam\ottott}\textfont\bffam=\ottobf%
\scriptfont\bffam=\sixbf\scriptscriptfont\bffam=\fivebf%
\def\bf{\fam\bffam\ottobf}\tt\ttglue=.5em plus.25em minus.15em%
\setbox\strutbox=\hbox{\vrule height7pt depth2pt width0pt}%
\normalbaselineskip=9pt\let\sc=\sixrm\normalbaselines\rm}

%
%
%
%
%
%
%

\global\newcount\numsec\global\newcount\numapp
\global\newcount\numfor\global\newcount\numfig\global\newcount\numsub
\numsec=0\numapp=0\numfig=1
\def\veroparagrafo{\number\numsec}\def\veraformula{\number\numfor}
\def\veraappendice{\number\numapp}\def\verasub{\number\numsub}
\def\verafigura{\number\numfig}

\def\section(#1,#2){\advance\numsec by 1\numfor=1\numsub=1%
\SIA p,#1,{\veroparagrafo} %
\write15{\string\Fp (#1){\secc(#1)}}%
\write16{ sec. #1 ==> \secc(#1)  }%
\hbox to \hsize{\titolo\hfill \number\numsec. #2\hfill%
\expandafter{\alato(sec. #1)}}\*}

\def\appendix(#1,#2){\advance\numapp by 1\numfor=1\numsub=1%
\SIA p,#1,{A\veraappendice} %
\write15{\string\Fp (#1){\secc(#1)}}%
\write16{ app. #1 ==> \secc(#1)  }%
\hbox to \hsize{\titolo\hfill Appendix A\number\numapp. #2\hfill%
\expandafter{\alato(app. #1)}}\*}

\def\senondefinito#1{\expandafter\ifx\csname#1\endcsname\relax}

\def\SIA #1,#2,#3 {\senondefinito{#1#2}%
\expandafter\xdef\csname #1#2\endcsname{#3}\else
\write16{???? ma #1#2 e' gia' stato definito !!!!} \fi}

\def \Fe(#1)#2{\SIA fe,#1,#2 }
\def \Fp(#1)#2{\SIA fp,#1,#2 }
\def \Fg(#1)#2{\SIA fg,#1,#2 }

\def\etichetta(#1){(\veroparagrafo.\veraformula)%
\SIA e,#1,(\veroparagrafo.\veraformula) %
\global\advance\numfor by 1%
\write15{\string\Fe (#1){\equ(#1)}}%
\write16{ EQ #1 ==> \equ(#1)  }}

\def\etichettaa(#1){(A\veraappendice.\veraformula)%
\SIA e,#1,(A\veraappendice.\veraformula) %
\global\advance\numfor by 1%
\write15{\string\Fe (#1){\equ(#1)}}%
\write16{ EQ #1 ==> \equ(#1) }}

\def\getichetta(#1){Fig. \verafigura%
\SIA g,#1,{\verafigura} %
\global\advance\numfig by 1%
\write15{\string\Fg (#1){\graf(#1)}}%
\write16{ Fig. #1 ==> \graf(#1) }}

\def\etichettap(#1){\veroparagrafo.\verasub%
\SIA p,#1,{\veroparagrafo.\verasub} %
\global\advance\numsub by 1%
\write15{\string\Fp (#1){\secc(#1)}}%
\write16{ par #1 ==> \secc(#1)  }}

\def\etichettapa(#1){A\veraappendice.\verasub%
\SIA p,#1,{A\veraappendice.\verasub} %
\global\advance\numsub by 1%
\write15{\string\Fp (#1){\secc(#1)}}%
\write16{ par #1 ==> \secc(#1)  }}

\def\Eq(#1){\eqno{\etichetta(#1)\alato(#1)}}
\def\eq(#1){\etichetta(#1)\alato(#1)}
\def\Eqa(#1){\eqno{\etichettaa(#1)\alato(#1)}}
\def\eqa(#1){\etichettaa(#1)\alato(#1)}
\def\eqg(#1){\getichetta(#1)\alato(fig. #1)}
\def\sub(#1){\0\palato(p. #1){\bf \etichettap(#1).}}
\def\asub(#1){\0\palato(p. #1){\bf \etichettapa(#1).}}

\def\equv(#1){\senondefinito{fe#1}$\clubsuit$#1%
\write16{eq. #1 non e' (ancora) definita}%
\else\csname fe#1\endcsname\fi}
\def\grafv(#1){\senondefinito{fg#1}$\clubsuit$#1%
\write16{fig. #1 non e' (ancora) definito}%
\else\csname fg#1\endcsname\fi}
\def\secv(#1){\senondefinito{fp#1}$\clubsuit$#1%
\write16{par. #1 non e' (ancora) definito}%
\else\csname fp#1\endcsname\fi}

\def\equ(#1){\senondefinito{e#1}\equv(#1)\else\csname e#1\endcsname\fi}
\def\graf(#1){\senondefinito{g#1}\grafv(#1)\else\csname g#1\endcsname\fi}
\def\secc(#1){\senondefinito{p#1}\secv(#1)\else\csname p#1\endcsname\fi}
\def\sec(#1){{\S\secc(#1)}}

\def\BOZZA{\bz=1
\def\alato(##1){\rlap{\kern-\hsize\kern-1.2truecm{$\scriptstyle##1$}}}
\def\palato(##1){\rlap{\kern-1.2truecm{$\scriptstyle##1$}}}
}

\def\alato(#1){}
\def\galato(#1){}
\def\palato(#1){}


{\count255=\time\divide\count255 by 60 \xdef\hourmin{\number\count255}
        \multiply\count255 by-60\advance\count255 by\time
   \xdef\hourmin{\hourmin:\ifnum\count255<10 0\fi\the\count255}}

\def\oramin{\hourmin }

\def\data{\number\day/\ifcase\month\or gennaio \or febbraio \or marzo \or
aprile \or maggio \or giugno \or luglio \or agosto \or settembre
\or ottobre \or novembre \or dicembre \fi/\number\year;\ \oramin}
\setbox200\hbox{$\scriptscriptstyle \data $}
\footline={\rlap{\hbox{\copy200}}\tenrm\hss \number\pageno\hss}

%
%
%
%
%
%
\newdimen\xshift \newdimen\xwidth \newdimen\yshift

\def\insertplot#1#2#3#4#5{\par%
\xwidth=#1 \xshift=\hsize \advance\xshift by-\xwidth \divide\xshift by 2%
\yshift=#2 \divide\yshift by 2%
\line{\hskip\xshift \vbox to #2{\vfil%
#3 \includegraphics{#4.ps}}\hfill \raise\yshift\hbox{#5}}}

\def\eqfig#1#2#3#4#5{ \par\xwidth=#1
\xshift=\hsize \advance\xshift by-\xwidth \divide\xshift by 2
\yshift=#2 \divide\yshift by 2 \line{\hglue\xshift \vbox to #2{\vfil #3
\includegraphics{#4.ps} }\hfill\raise\yshift\hbox{#5}}}

\def\8{\write13}


\let\a=\alpha \let\b=\beta  \let\g=\gamma     \let\d=\delta  \let\e=\varepsilon
\let\z=\zeta     \let\th=\theta    \let\l=\lambda
\let\m=\mu    \let\n=\nu                  \let\r=\rho
\let\s=\sigma \let\t=\tau        
   \let\o=\omega 
\let\G=\Gamma      \let\L=\Lambda  
    \let\Si=\Sigma \let\F=\Phi      
\let\O=\Omega 
\let\lg=\langle
\let\rg=\rangle

\def\\{\hfill\break} \let\==\equiv

\let\io=\infty 

\let\0=\noindent

\def\ie{\hbox{\it i.e.\ }}\def\eg{\hbox{\it e.g.\ }}
\let\dpr=\partial 

\def\tende#1{\,\vtop{\ialign{##\crcr\rightarrowfill\crcr
 \noalign{\kern-1pt\nointerlineskip}
 \hskip3.pt${\scriptstyle #1}$\hskip3.pt\crcr}}\,}
\def\otto{\,{\kern-1.truept\leftarrow\kern-5.truept\to\kern-1.truept}\,}

\def\PP{{\cal P}}\def\EE{{\cal E}}\def\MM{{\cal M}} \def\VV{{\cal V}}
\def\FF{{\cal F}}\def\HH{{\cal H}}
\def\BB{{\cal B}} 
\def\RR{{\cal R}}\def\LL{{\cal L}} \def\OO{{\cal O}}
\def\GG{{\cal G}} \def\SS{{\cal S}}
 
\def\T#1{{#1_{\kern-3pt\lower7pt\hbox{$\widetilde{}$}}\kern3pt}}
\def\VVV#1{{\underline #1}_{\kern-3pt
\lower7pt\hbox{$\widetilde{}$}}\kern3pt\,}
\def\W#1{#1_{\kern-3pt\lower7.5pt\hbox{$\widetilde{}$}}\kern2pt\,}

\def\indica{\leaders \hbox to 0.5cm{\hss.\hss}\hfill}
\def\guida{\leaders\hbox to 1em{\hss.\hss}\hfill}

\def\V#1{{\bf #1}}
\def\aa{{\V \a}}
\def\nn{{\V \n}}

\mathchardef\aa = "050B
\mathchardef\bb = "050C
\mathchardef\xxx= "0518
\mathchardef\zz = "0510
\mathchardef\oo = "0521
\mathchardef\ll = "0515
\mathchardef\mmm= "0516
\mathchardef\Dp = "0540
\mathchardef\H  = "0548
\mathchardef\FFF= "0546
\mathchardef\ppp= "0570
\mathchardef\nn = "0517
\mathchardef\pps= "0520
\mathchardef\XXX= "0504
\mathchardef\FFF= "0508
\mathchardef\tth= "0512

\def\us{\underline \s}

\def\der{{\rm d}}

\def\lft{\left}
\def\rgt{\right}

\def\la{\left\langle}
\def\ra{\right\rangle}

\def\qed{\raise1pt\hbox{\vrule height5pt width5pt depth0pt}}

\def\indic{\hbox{\raise-2pt \hbox{\indbf 1}}}

\def\RRR{\hbox{\msytw R}} 
 \def\CCC{\hbox{\msytw C}}
 
\def\NNN{\hbox{\msytw N}} 
 \def\ZZZ{\hbox{\msytw Z}}
 \def\zzz{\hbox{\msytwww Z}}

\newcount\mgnf  
\mgnf=0

\ifnum\mgnf=0
\def\openone{\leavevmode\hbox{\ninerm 1\kern-3.3pt\tenrm1}}%
\def\*{\vglue0.3truecm}\fi
\ifnum\mgnf=1
\def\openone{\leavevmode\hbox{\ninerm 1\kern-3.63pt\tenrm1}}%
\def\*{\vglue0.5truecm}\fi

\newcount\tipobib\newcount\bz\bz=0\newcount\aux\aux=1
\newdimen\bibskip\newdimen\maxit\maxit=0pt


\tipobib=0
\def\9#1{\ifnum\aux=1#1\else\relax\fi}

\newwrite\bib
\immediate\openout\bib=\jobname.bib
\global\newcount\bibex
\bibex=0
\def\verabib{\number\bibex}

\ifnum\tipobib=0
\def\cita#1{\expandafter\ifx\csname c#1\endcsname\relax
\hbox{$\clubsuit$}#1\write16{Manca #1 !}%
\else\csname c#1\endcsname\fi}
\def\rife#1#2#3{\immediate\write\bib{\string\raf{#2}{#3}{#1}}
\immediate\write15{\string\C(#1){[#2]}}
\setbox199=\hbox{#2}\ifnum\maxit < \wd199 \maxit=\wd199\fi}
\else
\def\cita#1{%
\expandafter\ifx\csname d#1\endcsname\relax%
\expandafter\ifx\csname c#1\endcsname\relax%
\hbox{$\clubsuit$}#1\write16{Manca #1 !}%
\else\probib(ref. numero )(#1)%
\csname c#1\endcsname%
\fi\else\csname d#1\endcsname\fi}%
\def\rife#1#2#3{\immediate\write15{\string\Cp(#1){%
\string\immediate\string\write\string\bib{\string\string\string\raf%
{\string\verabib}{#3}{#1}}%
\string\Cn(#1){[\string\verabib]}%
\string\CCc(#1)%
}}}%
\fi

\def\Cn(#1)#2{\expandafter\xdef\csname d#1\endcsname{#2}}
\def\CCc(#1){\csname d#1\endcsname}
\def\probib(#1)(#2){\global\advance\bibex+1%
\9{\immediate\write16{#1\verabib => #2}}%
}

\def\C(#1)#2{\SIA c,#1,{#2}}
\def\Cp(#1)#2{\SIAnx c,#1,{#2}}

\def\SIAnx #1,#2,#3 {\senondefinito{#1#2}%
\expandafter\def\csname#1#2\endcsname{#3}\else%
\write16{???? ma #1,#2 e' gia' stato definito !!!!}\fi}

\bibskip=10truept
\def\hboxto{\hbox to}

\catcode`\{=12\catcode`\}=12
\catcode`\<=1\catcode`\>=2
\immediate\write\bib<
        \string\halign{\string\hboxto \string\maxit%
        {\string #\string\hfill}&%
        \string\vtop{\string\parindent=0pt\string\advance\string\hsize%
        by -1.55truecm%
        \string#\string\vskip \bibskip
        }\string\cr%
>
\catcode`\{=1\catcode`\}=2
\catcode`\<=12\catcode`\>=12

\def\raf#1#2#3{\ifnum \bz=0 [#1]&#2 \cr\else
\llap{${}_{\rm #3}$}[#1]&#2\cr\fi}

\newread\bibin

\catcode`\{=12\catcode`\}=12
\catcode`\<=1\catcode`\>=2
\def\chiudibib<
\catcode`\{=12\catcode`\}=12
\catcode`\<=1\catcode`\>=2
\immediate\write\bib<}>
\catcode`\{=1\catcode`\}=2
\catcode`\<=12\catcode`\>=12
>
\catcode`\{=1\catcode`\}=2
\catcode`\<=12\catcode`\>=12

\def\makebiblio{
\ifnum\tipobib=0
\advance \maxit by 10pt
\else
\maxit=1.truecm
\fi
\chiudibib
\immediate \closeout\bib
\openin\bibin=\jobname.bib
\ifeof\bibin\relax\else
\raggedbottom
\input \jobname.bib
\fi
}

\openin13=#1.aux \ifeof13 \relax \else
\input #1.aux \closein13\fi
\openin14=\jobname.aux \ifeof14 \relax \else
\input \jobname.aux \closein14 \fi
\immediate\openout15=\jobname.aux

\def\biblio{\*\*\centerline{\titolo References}\*\nobreak\makebiblio}
\def\defi{{\buildrel def \over =}}


\ifnum\mgnf=0
   \magnification=\magstep0
   \hsize=16.5truecm\vsize=21.5truecm
   \parindent=0.3cm\baselineskip=0.45cm\fi
\ifnum\mgnf=1
   \magnification=\magstep1\hoffset=0.truecm
   \hsize=16truecm\vsize=21.5truecm
   \baselineskip=18truept plus0.1pt minus0.1pt \parindent=0.9truecm
   \lineskip=0.5truecm\lineskiplimit=0.1pt      \parskip=0.1pt plus1pt\fi

 
\voffset1.5truecm
\hoffset1.2truecm

 \hsize=13.5truecm\vsize=19.5truecm
   \baselineskip=16truept plus0.1pt minus0.1pt \parindent=0.9truecm
   \lineskip=0.5truecm\lineskiplimit=0.1pt      \parskip=0.1pt

\footline{\hfill\the\pageno\hfill}

\null\vskip.6truecm

\centerline{\titolone  Analyticity of the SRB measure}\smallskip
\centerline{\titolone for a class of simple Anosov flows}

\vskip1.truecm

\centerline{{\titolo
A. Amaricci$^{1}$, F. Bonetto$^{2}$ and P. Falco$^{3}$ }}
\vskip.2truecm

\centerline{{}$^{1}$ Dipartimento di Fisica, Universit\`a di Roma ``Tor Vergata''; and }
\centerline{Laboratoire de Physique des Solides, Universit\'e de Paris-Sud XI, Orsay}
\vskip.1truecm
\centerline{{}$^{2}$ School of Mathematics, Georgia Institute of Technology, Atlanta }
\vskip.1truecm
\centerline{{}$^{3}$ Dipartimento di Matematica, Universit\`a di Roma ``Tor Vergata''; and }
\centerline{Junior Research Fellow at ESI, Vienna, Austria}
\vskip1.truecm

\line{\vtop{
\line{\hskip1.5truecm\vbox{\advance \hsize by -3.1 truecm
\0{\cs Abstract.}
{\it We consider perturbations of the Hamiltonian flow associated 
 with the geodesic flow on a surface of constant negative curvature. We 
prove that, under a small perturbation, not necessarily of Hamiltonian
character, the SRB measure associated to the flow exists and is
analytic in the strength of the perturbation. An explicit example of
``thermostatted'' dissipative dynamics is considered.}} \hfill} }}

\vskip1.truecm
 \section(I,Introduction)
 
 \sub(1) In recent time, much effort has been devoted to the analysis
 of hyperbolic systems, in part due to the {\it Chaotic Hypothesis},
 introduced ten years ago in \cita{GC}, which states that a many
 particles systems in a nonequilibrium stationary state behave as a
 uniformly hyperbolic dynamical system (Anosov or more generally Axiom
 A system), at least for the purpose of evaluating macroscopic
 observables.  This hypothesis can be seen as a generalization of the
 ergodic hypothesis to out of equilibrium systems, at least for
 systems in a stationary states.  Although it is very hard to prove
 uniform hyperbolicity for realistic model systems, ideas connected
 to the Chaotic hypothesis have played an important role in analyzing
 the results of numerical or real experiments.
 
 Several results have been obtained in this sense, among which the
 Gallavotti-Cohen {\it Fluctuation Theorem} (FT), a result concerning
 the large deviation functional of the phase space contraction rate
 (often identified with the {\it entropy production rate}), that
 extend the fluctuation-dissipation relation to systems in a
 non-equilibrium stationary state.  The FT was proved rigorously in
 \cita{G} for Anosov diffeomorphisms and then in \cita{Ge} for Anosov
 flows.  Furthermore several numerical tests have been conducted, using
 mathematical models of dissipative reversible systems and the chaotic
 hypothesis. 
 
 Most of the results quoted above are based on the existence of the
 Sinai-Ruelle-Bowen (SRB) measure. This existence was proved for a
 wide class of hyperbolic systems \cita{BR},\cita{S}. Unfortunately
 explicit expressions for the SRB measure are quite difficult to
 obtain and can be worked out only in particular cases, \eg Anosov
 Coupled Lattice Map \cita{BFG}, while most of the models used in the
 simulations are based on continuous time dynamics (hyperbolic flows).
 We observe that, in order to obtain results for nonequilibrium
 stationary systems, one can not consider the simplest examples of
 Anosov systems that, being volume preserving, are not dissipative.
 
 In this paper we explicitly construct the SRB measure for a family of
 Anosov flows that includes dissipative cases. The flows considered
 are perturbations of the geodesic flow on a surface of constant
 negative curvature. Such a flow can be seen as the flow generated by
 an Hamiltonian dynamics on the surface of unit energy.  We will
 mainly consider perturbation arising by adding a force to the
 Hamiltonian equations of motion. If the chosen force is conservative
 (\ie coming from a potential), then the system remain Hamiltonian and
 volume preserving so that the stationary measure is not singular with
 respect to the volume measure.  Otherwise, if the perturbation is non
 conservative, the system is expected to have an SRB measure singular
 with respect to the volume measure (dissipativity).  Many of the models
 used in the numerical works fall under this last category.
 
The paper is organized as follows. In section 2 we introduce the
systems we will consider and state the main results of the
paper. Section 3,4,5 contain the proof of these results. A
conclusive section gives comparison with known works and outlooks.
Finally the Appendices contain some technical computations.
\* \*

\section(mr,Model and main results)

\sub(M){\it The geodesic flow}. The complex upper half plane
$\CCC_+\defi\{z\in \CCC:{\rm Im}(z)>0\}$, endowed with the metric
$g=y^{-2}\pmatrix{1 & 0\cr 0 & 1}$, is called the {\it Lobachevskii
  plane}. The isometries of this plane are given by the real,
$2\times2$ matrices $h$ with $\det h>0$ where, if $z\in \CCC_+$, the
action of $h$ on $z$ is
$$
 zh\defi{h_{11}z+h_{21}\over h_{12}z + h_{22}}\in \CCC_+\;,
\quad{\rm for\ }
h\defi\pmatrix{h_{11}& h_{12}\cr
               h_{21}& h_{22}}\;.
$$
Observe that $h$ and $h'=\l h$, for $\l\not=0$, define the same
transformation so that such isometries are
naturally represented by the elements of $\rm PSL(2,\RRR)$. 

A compact surface can be constructed from the Lobachevskii plane
in the same way as the  torus can be obtained from the plane $\RRR^2$, 
under identification
of the points $(x,y)$ and $(x+m,y+n)$, for any $n,m\in \ZZZ$.
Indeed, for any Fuchsian subgroup $\G\subset {\rm PSL(2,\RRR)}$ (see \cita{P}
for a precise definition), we can consider the equivalence relation
generated by its action on $\CCC_+$,

$$z\sim z'\iff  \exists \g\in \G \,|\,z= z'\g\;.$$
The quotient set, indicated with $\Si=\CCC_+/\G$ is the 
most general compact analytic surface with constant negative curvature. 
If $\G$ is the
smallest possible Fuchsian subgroup, we obtain a surface of genus two
(2-torus); in the following we will mainly consider this surface
although our results apply to all above surfaces.

We will consider as {\it unperturbed dynamical system} the 
flow generated by the Hamiltonian
$$
H_0(x,y,p_x, p_y)\defi {y^2\over 2}\lft(p^2_x+p^2_y\rgt)\;\Eq(H)
$$
on the cotangent bundle $\MM\defi
T^*\Si$. For any given energy $\EE>0$,
the surface 
$$\MM_\EE\defi\big\{(x,y,p_x,p_y)\in \MM:H_0(x,y,p_x,p_y)\=\EE\big\}\;,$$ 
is a compact, invariant manifold. The geodesic flow on the surface $\Si$ 
can be seen in a natural way as an Hamiltonian flow generated by \equ(H) 
on $\MM_1$.

We want to add a conservative force to such a system. Given a
$\G$--periodic function $\{V(z), z\in \CCC_+\}$, we can consider the
new Hamiltonian function
$$
H_\e=H_0+\e V\Eq(pH)
$$
which generates the equations of motion
$$
\lft\{ \eqalign{ \dot x&=y^2p_x\;, \qquad \dot
p_x=-\e{\dpr V\over \dpr x}\;, \cr \dot y&=y^2p_y\;, 
\qquad \dot p_y=-y(p_x^2+p_y^2)-\e{\dpr V\over \dpr y}\;.}\rgt. \Eq(sd)
$$
We can then add a non-conservative force to our system. Since it
has to be {\it covariant} w.r.t. the group of transformations in $\G$,
the simplest such a field can be written in terms of the {\it
  automorphic} function, $\phi$, and the {\it anti-automorphic} one,
$\overline\phi$, of order one (see \cita{F}):
$$
E_x={\overline{\phi(z)}+\phi(z)\over2}\ , \qquad\quad
E_y={\overline{\phi(z)}-\phi(z)\over2i}\;.
$$
The defining property of $\phi$ and $\overline\phi$ is that,
calling $ j(z,h)\defi h_{12}z+h_{22}$, they satisfy the identity:
$$
\phi(z\g)=\phi(z)j^2(z,\g)\ ,\qquad \forall\ \g\in\G\;,
$$
so that, the ``potential difference''
between two points, $z,z_0\in \Si$,
$$
U(z)-U(z_0)\defi \int_{z_0}^z\!\der w\ \phi_1(w)+
\int_{z_0}^z\!\der \overline w\ \overline{\phi_1(w)}
$$
is well defined. Such a field is locally conservative, but it is not
the differential of a function. Hence the energy $H_\e$ computed along a
motion of \equ(sdbis)
asymptotically tends to increases. In order to maintain it constant, we
introduce a Gaussian thermostat, namely a momentum-dependent friction
of the form $\a(p)=p\cdot E/p^2$. Finally, the equations of motion for
the {\it perturbed flow} on $\MM^\e_\EE\defi\{(x,y,p_x,p_y)\in
\MM:H_\e(x,y,p_x,p_y)\=\EE\}$ are:
$$
\lft\{ 
\eqalign{ 
\dot x&=y^2 p_x\;, 
\qquad \dot
p_x=-\e{\dpr V\over \dpr x}+\e'\lft[E_x - \a(p) p_x\rgt]\;, \cr 
\dot y&=y^2p_y\;, \qquad 
\dot p_y=-y(p^2_x+p^2_y)-\e{\dpr V\over \dpr y}
+\e'\lft[E_y-\a(p) p_y\rgt]\;.}\rgt. \Eq(sdbis)
$$
where $\e'$ is the strength of the non-conservative field: since only
notational complication would arise from considering $\e\neq \e'$, in
the following, we will restrict ourselves to the case $\e=\e'$.  Under
the dynamics eq.\equ(sdbis) $H_\e$ is an integral of the motion.  \*

\sub(CC){\it Canonical coordinates.} A simpler representation of the
umperturbed dynamics was introduced in \cita{CEG}.  
We consider the {\it canonical transformation} from
$\MM\backslash\{H_0=0\}$ to $\GG \defi {\rm GL}(2,\RRR)/\G$
$$
(p_x,p_y,x,y)\longleftrightarrow
\pmatrix{p_1&q_2\cr -p_2&q_1}\defi g\;,
$$
defined by
$$
\left\{\eqalign{
& p_x+ip_y={i\over 2}{\rm det}^2(g)j^2\lft(i,g^{-1}\rgt)\cr
& x+iy=ig^{-1}\;.}\right. \Eq(D)
$$
This transforms the equations of motion \equ(sd) into those generated
by the new Hamiltonian (with slight abuse of notation, we still call
$H_\e$ and $V$ the Hamiltonian and the potential as function of the
matrix $g$)
$$
 H_\e(g)\defi{\det^2(g)\over 8}+\e V(g)\;.\Eq(nh)
$$
Clearly $H_\e$ is an analytic function of $g$.  Considering the
following matrices
$$
 \s^0\defi\pmatrix {1 & 0 \cr 0 & 1}\;,\qquad
 \s^3\defi\pmatrix {1 & 0 \cr 0 & -1}\;,\qquad
 \s^+\defi\pmatrix {0 & 1 \cr 0 & 0}\;,\qquad
 \s^-\defi\pmatrix {0 & 0 \cr 1 & 0}\;;\Eq(M)
$$
the Hamilton equation derived from
\equ(nh) reads
$$
\dot g= -{\det(g)\over4}g\s^3 +\e\s^x{\dpr V\over \dpr g} (g)\s^y \;, \Eq(He)
$$
for $\s^x\defi (\s^++\s^-)$ and $\s^y\defi (\s^+-\s^-)$.
The non-conservative equations of motion \equ(sdbis) reads
$$\eqalign{
\dot g 
&=-{\det(g)\over 4}g\s^3+\e\s^x{\dpr V\over \dpr g} (g)\s^y 
-\e c(g)g\s^y
\cr &
\defi {\det(g)\over 4}\lft[-g\s^3 + \e\FF(g)\rgt]\;,
}\Eq(Hebis)$$
where the function $c(g)$ is:
$$
c(g)={1\over2\det^2(g)}\left[{\phi(ig^{-1})\over{j^2(i,g^{-1})}}+
{\overline{\phi(ig^{-1})}\over{j^2(-i,g^{-1})}}\right]\;.
$$
This is a explicit example of a non-conservative system. 
\*
\sub(R1){\it Remark.}
Our techniques can be extended to a more general case. Given an
Hamiltonian 
$$
 H_\e(g)\defi H_0(g)+\e V(g)\Eq(ham)
$$
like in eq.\equ(nh) we can consider any analytic vector field $\VV_\e$
on $\MM$, $\e$-close to the Hamiltonian vector field generated by
$H_0$ and tangent to the level surfaces of $H_\e$. Clearly the flow
generated by such a vector field preserve $H_\e$ and the following
results hold in this more general situation.  \*

\sub(T1){\it The conjugation}. Let $\F_t:\GG_\EE\to\GG_\EE $ and 
$\F^\e_t:\GG_\EE^\e\to\GG_\EE^\e$ be the flows
generated by the Hamiltonian $H_0$ and by the dissipative system in
eq.\equ(Hebis), respectively. As a first step we want to prove 
that this two flows can be conjugated by a change of coordinate.
Differently from the case of Anosov diffeomorphisms, \cita{GBG}, this
is not enough to map $\F_t$ into $\F^\e_t$, but
a local rescaling of time is also required.  
The precise details are given in the
following theorem. To state it we need some notations:
$$\eqalign{
\GG_\EE=\{g\in \GG\,|\, H_0(g)=\EE\}\;,&\qquad
\GG_{>\EE}=\{g\in \GG\,|\, H_0(g)>\EE\}\;,\cr
\GG^\e_\EE=\{g\in \GG\,|\, H_\e(g)=\EE\}\;,&\qquad
\GG^\e_{>\EE}=\{g\in \GG\,|\, H_\e(g)>\EE\}\;.}$$

{\bf\0Theorem 1. Conjugation.}  {\it Given $\EE>0$, there exists an
  $\bar\e>0$ such that, for any $\e:|\e|\le \bar\e$ there are
  functions $h_\e:\GG_{>\EE}\to\GG^\e_{>\EE}$, and
  $\t_\e:\GG_{>\EE}\to\RRR$, H\"older continuous in $g$ and analytic
  in $\e$, such that
$$
h_\e\circ \F_t =\F^\e_{T^\e_t}\circ h_\e\;,
\qquad
{\rm for \ \ }
T^\e_t\defi\int_0^t\!\!\der s\
\lft(\t_\e\circ \F_s\rgt)\;. \Eq(conj)
$$
Furthermore, $H_0\=H_\e\circ h_\e$, so that
$h_\e(\GG_{\EE})=\GG^\e_{\EE}$.}  
\*
\0The proof, given in section 3,  is based on the hyperbolicity of the
unperturbed flow, which is discussed in the next section.

The function $h_\e$ is the {\it space conjugation}, while $\t_\e$ is the
{\it time conjugation}.  Even if $h_\e$ conjugate the flow from $\GG_{\EE}$
to $\GG^\e_{\EE}$, the existence of a conjugation from whole the $\GG$
to itself can not be uniform in $\e$.  Indeed, fixed $\e$, if
$\EE<\e \sup_g V(g)$
the topology of $\GG^\e_\EE$ is different from that of 
$\GG_\EE$, and no conjugation is possible. 

\* \*

\sub(hh){\it Hyperbolicity}.
If the tangent space $T_g\GG^\e_\EE$ can be splitted into three
continuous, $\Phi^\e$-covariant, one-dimensional, linear subspaces:
$$
 T_g\GG^\e_\EE=E_g^+ \oplus E_g^- \oplus E_g^3\Eq(wit)
$$
such that $E_g^3$ is parallel to the flow; and 
if there are constants $c,\l>0$ such that
$$
\eqalign{
&\|\lft(T_g\Phi^\e_t\rgt)w\|\leq ce^{-\l t}\|w\|\quad {\rm for}\ \  w\in E_g^-,\ t\geq0\cr
&\|\lft(T_g\Phi^\e_t\rgt)w\|\leq ce^{\l t}\|w\|\quad {\rm for}\ \  w\in E_g^+,\ t\leq0\;,\cr}
\Eq(105)
$$
then the flow $\Phi^\e$ is {\it hyperbolic} on $\GG_\EE^\e$. Moreover
$E_g^+$, $E_g^-$ and $E_g^3$ are called the {\it unstable}, {\it stable}
and {\it neutral} subspace, respectively.

The umperturbed flow, $\Phi$, is hyperbolic on $\GG_\EE$, for every
$\EE>0$.  The solution of \equ(He) is explicitly given by:
$$\Phi_t(g)\defi
ge^{-(\det(g)/4)t\s_3}\ {\rm mod}\ \G.\Eq(F)
$$
and it is clear that $E^\a_g$ is generated by 
$g\s^\a$, for $\a=\pm,3$ and $\l=\sqrt{2 \EE}$. 

The four curves 
$$
\F^\a_\z(g)\defi ge^{-\z\s^\a}\ {\rm mod}\ \G \qquad{\rm for \ \
}\a=3,0,\pm\Eq(G)
$$
are the integral manifold of the vector fields $w^a(g)\defi-g\s^a$,
for $a=0,\pm,3$. We remark that $\F_t\= \F^3_{t \det(g)/4}$
and that $\F^0$ is orthogonal  to  $\GG_\EE$.

Calling $\l^\pm(g)=\pm\det(g)/2=\pm \sqrt{2H_0(g)}$ and $\l^3\=0$ the
{\it Ljapunov exponents} of $\Phi_t$, and using that the commutation
relation among the matrices $\{\s^i\}_{i=0,3,\pm}$ are
$$
[\s^3,\s^+]=2\s^+ \qquad [\s^3,\s^-]=-2\s^- \qquad
[\s^+,\s^-]=\s^3 \;, \Eq(108)
$$
we obtain that:
$$
\F_t\circ \F^\a_\z=\F^\a_{\z \exp\{t\l^\a(g) \}}\circ\Phi_t\;.
\Eq(104)
$$
\* \*

{\bf\0Theorem 2. Hyperbolicity.}  {\it For any energy $\EE>0$,
  there exists $\bar\e>0$ such that, for any $|\e|\le \bar\e$ the flow
  $\F^\e$ on $\GG_\EE^\e$ is hyperbolic.  In particular, there exist
  vector fields $\{w_\e^\a\}_{\a=0,\pm}$ and functions
  $\{\l_\e^\a\}_{\a=0,\pm}$ such that
$$
 T\Phi^\e_t\ w_\e^\a
 =
 \exp\lft\{\int_0^t\!\der s\
 \big(\l^\a_\e\circ \Phi^\e_s\big)\rgt\} 
\lft(w^\a_\e\circ \Phi^\e_t\rgt)\;, \qquad
{\rm for \ }\a=0,\pm\;. \Eq(301)
$$
Furthermore, $\{w_\e^\a\circ h_\e\}_{\a=0,\pm}$ and $\{\l^\a_\e\circ
h_\e\}_{\a=0,\pm}$, are analytic in $\e$, and H\"older continuous in
$g$.}  
\* 
\0Notwithstanding we called the conjugation a change of
variables, since it is not differentiable -but only H\"older-continuous-
this theorem is {\it not} a direct consequence of theorem 1. The fact
that $\{\l_\e^\a\circ h_\e\}_{\a=0,\pm}$, rather than
$\{\l_\e^\a\}_{\a=0,\pm}$, are analytic in $\e$, will be important for
the construction of the SRB measure.  \* \*

\sub(srbd){\it SRB distribution.}
For any energy $\EE$ we can define the  SRB measure on $\GG_\EE^\e$:
$$
\m^{\e}(\OO)\defi
\lim_{T\to\io}{1\over T}\int_0^T\!\der t\ \big(\OO\circ\F^\e_t\big)(g)\;,\Eq(srb)
$$
provided that such a limit exists and is constant Lebesgue-almost
everywhere in $g$ for every continuous function $\O$.  Such a measure
exists, is unique and ergodic, if the dynamical system is Anosov, i.e. 
it is  hyperbolic in the whole $\GG_\EE^\e$.

The flow  $\Phi$, besides being  Anosov, is topologically mixing: the stable 
and the unstable manifold are dense $\GG_\EE$;  since it is also
Hamiltonian, it is easy to prove that its SRB measure is the Lebesgue
measure.  

Regarding  $\F^\e$, uniform hyperbolicity  was established in
Theorem 2, while the topological mixing is a direct consequence of the existence 
of the conjugation.
\* \*
{\bf\0Theorem 3. Analyticity of the SRB measure.} {\it 
Given $\EE>0$, there exists $\bar\e>0$, such that,
for any $|\e|<\bar\e$ the SRB measure $\m^\e$ is analytic in $\e$: 
\ie, for any  analytic
$\OO:\GG\to\RRR$, the mean value $\m^{\e}(\OO)$ is analytic in $\e$.} 
\* 
\0This is our  main result. The proof will consist
in an explicit construction of the SRB measure. 

To summarize, for any energy $\EE>0$, and $\e$ small enough,
we have constructed an hyperbolic structure 
and the corresponding SRB measure 
on each one of the leaves $\{\GG^\e_{\EE'}\}_{\EE'\ge\EE}$
in $\GG^\e_{\ge\EE}$. The set $\{\m^\e_{\EE'}\}_{\EE'>\EE}$
is an invariant measure on $\GG^\e_{>\EE}$.
 
\* 
\*
  \section(CAF,Proof of Theorem 1.)

\sub(PCC){\it Directional derivatives}.
For any smooth $f$ on $\GG$
we define the {\it directional derivative} along the curves 
$\{\F^\a\}_{\a=0,\pm,3}$, as:
$$
 \big(\LL_\a f\big)(g)
\defi
 {d\big(f\circ \F^\a_\z\big)\over d\z}\Big|_{\z=0}(g)\;.\Eq(Hbis)
$$
These derivatives satisfy the relation
$\big(\LL_\a w^3\big)
 -\big(\LL_3 w^\a\big)=\l^\a w^\a\;.$
Since the stable, unstable and neutral directions are tangent to
$\GG_\EE$, whereas $w^0$ is transversal to it, we have
$$\eqalign{
 \big(\LL_\a H_0\big)(g)
&\equiv0\qquad {\rm for} \quad\a=3,\pm \quad g\in\GG_\EE\cr
 \big(\LL_0 H_0\big)(g)
&\neq0\qquad{\rm for}\quad  g\in\GG_\EE\;. }\Eq(107)$$
Given $\g<1$ and a function $f$ on $\GG$,
we also define the  {\it directional H\"older derivative}
along $\{\F^\a\}_{\a=0,\pm,3}$ as
$$
  \big(\LL^\g_\a f\big)(g)
\defi
  \sup_{\z:0<|\z|\le 1}{\lft|\big(f\circ \F^\a_\z\big)(g)-f(g)\right|\over{|\z|^\g}}\;,\Eq(I)
$$
if the supremum is finite.
\*
\*

\sub(PC){\it Construction of the Conjugation}. 
In order to find a solution of
\equ(conj), let us differentiate it w.r.t. $t$, for $t=0$:
$$
\big(\LL_3 h_\e\big)(g)={\big(\det\circ h_\e\big)(g)\over
\det(g)}\t_\e(g)\Big[w^3\circ h_\e+\e \FF\circ h_\e\Big](g)\;.\Eq(cc)
$$
We will look for 
a solution $h_\e$ and $\t_\e$ of the form
$$\eqalign{
&h_\e(g)=g+\sum_{\a=0,\pm,3}\d h^\a_\e(g) w^\a(g)=
\sum_{\a=0,\pm,3}\big[\d_{0,\a}+\d h^\a_\e(g)\big] w^\a(g) \cr &
\t_\e=1+\d\t_\e} \Eq(anz1)
$$
Projecting along the directions $\{w^\a(g)\}_{\a=0,\pm,3}$ and using
the identity following \equ(H), yields
(see the Appendix for the explicit developments):
$$\eqalign{
 \big(\LL_3 \d h_\e^\a\big)(g)-\l^\a\d h_\e^\a(g)
 =&
 \e\FF^\a(g)+
 \RR^\a_\e\big(\d h^0_\e,\d h^3_\e,\d h^+_\e,\d h^-_\e,\d\t_\e\big)
 \cr&
 +\d_{\a,3}\big(\d\t_\e(g)-2\d h^0_\e(g)\big)
 }\Eq(202)
$$
where $\d_{\a,\b}$ is the Kronecker symbol.
In the r.h.s. member of \equ(202), $\{\FF^\a:\GG\to \RRR\}_{\a=0,\pm,3}$
are analytic function of $g$, depending neither from $\d h_e$, nor from $\d\t_\e$;
while 
$\{\RR^\a_\e:\RRR^5\to \RRR\}_{\a=0,\pm,3}$ are certain analytic function of the form 
$$
\RR^\a_\e(f_1,f_2,f_3,f_4,f_5)=
\e\sum_{i=1}^5 C^\RR_{\a,i} f_i +{\rm O}(f^2)\Eq(DD)
$$
for suitable constants $\{C^\RR_{\a,i}\}_{i=1,\ldots,5\atop \a=0,\pm,3}$,
and ${\rm O}(f^2)$ of order 0 in $\e$.
The last term in \equ(202) is linear in $\d h_\e^0$, and in $\d\t_\e$,
but we singled it out  because it is order 0 in $\e$.  
\* \*

\sub(sol){\it Implicit solution}. We can implicitly solve
\equ(202). For every continuous $f:\GG\to\RRR$, it is
possible to invert the operators $\{\LL_3-\l^\b\}_{\b=\pm}$:
$$
\left(\LL_3 - \l^\b\right)^{-1}f 
 =
\int_{{\rm sgn}(\l^\b)\io}^0\!\der t\ 
e^{(\LL_3-\l^\b)t}f
 =
 \int_{{\rm sgn}(\l^\b)\io}^0\!\der t\
e^{-\l^\b t}\big(f\circ\Phi_t\big)\;,
\qquad \b=\pm\Eq(O)
$$
where the exponential decaying factor guarantees convergence.

The implicit solution for the stable and the unstable components of the
conjugation are then:
$$
\eqalign{
  \d h_\e^\b
 &=
  \int_{{\rm sgn}(\b)\infty}^0\! \der t\
  e^{-\l^\b t}
  \big(\RR^\b_\e\circ \Phi_t\big)+\e
  \int_{{\rm sgn}(\b)\infty}^0\! \der t\
  e^{-\l^\b t}
  \big(\FF^\b_\e\circ \Phi_t\big)\;,
\qquad \b=\pm\;,}\Eq(205)
$$
for $\RR^\b_\e\circ \Phi_t\defi\RR^\a_\e\lft(\{\d h^\a_\e\circ 
\Phi_t\}_{\a=0,\pm,3},\d\t_\e\circ \Phi_t\rgt)$.
The equation for $\d h_\e^3$ cannot be solved in the same way since $\l^3\=0$.
Nonetheless, we can choose $\t_\e$ so that the r.h.s. member of \equ(202), for $\a=3$,
is identically zero:
$$\eqalign{
 \d \t_\e=
 &2\d h^0_\e
 -\e\FF^3 
 -\RR^3_\e\big(\d h^0_\e,
 0,\d h^+_\e,\d h^-_\e,\d\t_\e\big)\;.}\Eq(206)
$$
Since also $\l^0\=0$, a similar problem occurs for the equation
corresponding to $\d h_\e^0$: in this case, it is possible 
to obtain an equation for $\d h_\e^0$ using that
$H_\e\circ h_\e=H_0$.
Considering the transversality condition \equ(107) and the
implicit equations for the level surfaces, one can solve \equ(202)
in terms of $\d h^0_\e$ only, obtaining:
$$
\eqalign{
 \d h_\e^0
&= -{1\over \LL_0 H_0} \lft[ H_0\circ h_\e -
H_0-\sum_\a\big(\LL_\a H_0\big)\cdot\d h_\e^\a +\e V\circ
h_\e\rgt]\cr &
\defi
-\e{V \over \LL_0H_0}-\OO \big(\d h^0_\e,\d h^3_\e,\d h^+_\e,\d h^-_\e,\d
\t_\e\big)\;, }\Eq(208)
$$
where $\OO$ can be written as in \equ(DD), for certain other constants
$\{C^\OO_{\a,i}\}_{i=1,\ldots,5\atop \a=0,\pm,3}$. 
The fact that $w^0$ is orthogonal to the
level surfaces of $H_0$, (see \equ(107)) guarantees that this
expression is well defined for any $g\in \GG_\EE$ and $\e$ small
enough. \* \*

\sub(EC){\it Existence of the conjugation}.
Observe that the equations 
\equ(205), \equ(206) and \equ(208) can be naturally seen as
defining a function $f\defi\{f^\a:\GG_\EE\to \RRR^4\}_{\a=0,\pm,3}$ 
for $f^0=\d h^0_\e$, 
$f^\pm=\d h^+_\e$ and  $f^3=\d\t_\e$. 
We will look for the solutions of the above equations
in the Banach space, $\BB$, defined by the norm $\|f\|_\g \defi\max_\a \|f^\a\|_\g$,
for
$$
 \|f^\a\|_\g \defi\|f^\a\|+ \sum_{\b=\pm}\|\LL_\b^\g f^\a\|
 +\sum_{\b=3,0}\|\LL_\b f^\a\| \;,
$$
where, if $u:\GG\longrightarrow \RRR$, then
$\|u\|\defi\sup_{g\in\GG}|u(g)|$.

The equation for the conjugation is given in terms of the operator

$$
\big(L f\big)^\a\ 
\defi
\left\{\eqalign
{& (\LL_3 -\l^\a) f^\a
\quad {\rm if}\ \a=\pm\cr
 & f^\a\phantom{********}
\quad {\rm if}\ \a=0,3}\right.\;,$$
and the function
$$S^\a_\e \big(f\big)
\defi\cases{
 \e\FF^\a+
 \RR^\a_\e\big(f^0,0,f^+,f^-,f^3\big) & for $\a=\pm$\cr
-\e(\LL_0 H_0)^{-1}\cdot V+\OO \big(f^0,0,f^+,f^-,f^3\big) & for $\a=0$\cr
-\e\lft[(\LL_0 H_0)^{-1}\cdot 2V+\FF^3\rgt] 
 -\Big(2\OO +\RR^3_\e\Big)\big(f^0,0,f^+,f^-,f^3\big) & for $\a=3$}\Eq(Q)$$
\*\*
{\bf\0Lemma 1.}
{\it There exists $\bar \e>0$ such that, for any $|\e|\le \bar\e$,
the equation 
$$
 Lf=S_\e (f)\Eq(th1)
$$
has unique solution in the ball of $\BB$ with radius $|\e| C$, for a suitable $C$.
Such a solution is analytic in
$\e$.}
\*

{\bf\0Proof.} 
We first bound the norm of  $L^{-1}$.
From \equ(104) it follows that 
$$\eqalign{
 \sup_{|\z|>0}{|f\circ\F_t\circ \F_\z^\a-f\circ\F_t|\over |\z|^\g}
 &= e^{\g t\l^\a}\sup_{\z>0}
{\big|\big(f\circ \F_{\z\exp\{t\l^\a\}}^\a-f\big)\circ\F_t\big|
\over |\z|^\g \exp\{\g t \l^\a\}} 
\cr&
\le e^{\g t \l^\a}\Big(\|\LL^\g_\a
f\|+2\|f\|\Big)\;;
}$$
from this, it is easy to get the bound $\|L^{-1}\|_\g\le 5/\l_+(1-\g)$.

We choose $C \ge \|L^{-1}\|\cdot\max
\{1,\ 4\|\FF\|_\g,\  4\|(\LL_0 H_0)^{-1} V\|_\g\}$. 
From \equ(DD), there exists a $\g,\e$-independent constant
$C_0>1$ such that, for any $f,\tilde f$ in the 
ball $\BB_{\e}\defi\{f\in \BB:\|f\|_\g\le |\e| C\}$,  
$$
\|\OO(f)- \OO(\tilde f)\|_\g,\ \|\RR^\a_\e\big(f)-\RR^\a_\e\big(\tilde f)\|_\g
\le|\e| C_0\|f-\tilde f\|_\g \;. \Eq(b1)
$$
Indeed, it is possible to 
write $\OO(f)- \OO(\tilde f)=\sum_{j=1}^5(f_j-\tilde f_j)\int_0^1\!dt\ (\dpr_j\OO)\circ(tf+(1-t)\tilde f)$
and similarly for $\RR^\a$;
 furthermore, the H\"older derivative of a product
of functions 
is bounded by the product of the H\"older derivatives of each functions.
From \equ(b1) it follows
$$
 \|S^\a_\e(f)-S^\a_\e(\tilde f)\|_\g\le |\e| 3C_0 \|f-\tilde f\|_\g\;. \Eq(b2)
$$
By the choice of $C$ and using 
\equ(b2) for $\tilde f\=0$,  we have that, choosing 
$\bar \e=\l_+ {1-\g\over 60 C_0}$, $L^{-1}S_\e$ sends $\BB_{\e}$
into itself. Moreover \equ(b2) 
implies that the application $L^{-1}S_\e$ is a contraction
in $\BB_{\e}$ since, by the previous choice,
$\bar\e<\l_+{1-\g\over 20 C_0}$.
Since $\FF$ and  $V$ are analytic,
the solution of \equ(th1) is unique in $\BB_{\e}$ and is the
limit of  a sequence of functions which are analytic in
$\{\e\in\CCC:|\e|\le \bar \e\}$: by Vitali theorem
the solution is also analytic.
\hfill\qed\hskip1em\null
\*
\*
This Lemma concludes the proof of Theorem 1. 
\*\*

  \section(HPF, Proof of Theorem 2.)

\sub(UD){\it Unstable Direction}. The second step towards the construction
of an analytic SRB measure for the model considered in this paper,
is to obtain the perturbed unstable direction $w^+_\e(g)$ and
the associated Ljapunov exponent $\l^+_\e(g)$.
These quantities are both defined by \equ(301).

As expected from the general theory of the Anosov flows \cita{A}, the
unstable direction of the perturbed system $w_\e^+$ is {\it generically}
not analytic in $\e$. To construct the SRB measure we need
unstable direction computed in the conjugated
point $h_\e$, which we will see to be analytic in $\e$.

Calling $v^+_\e\defi w_\e^+\circ h_\e$ and 
$L_\e\defi\l^+_\e \circ h_\e \defi\l^+ + \d L_\e$, it is convenient to 
compute \equ(301) for time $t$ replaced by $T^\e_\t$ and 
position $h_\e(g)$ rather than $g$. Using also \equ(conj),
it follows:
$$
 \big(T_{h_\e}\Phi^\e_{T^\e_t} \big)\ v^+_\e
 =
 e^{\int_0^t\!\der s\
 \lft(\t_\e \circ \Phi_s\rgt)\lft(L^+_\e\circ \Phi_s\rgt)}\ 
\lft(v^+_\e\circ \Phi_t\rgt). \Eq(302) 
$$
\*
\*

\sub(CUD){\it Construction of the Unstable Direction}. Proceeding
as in the previous section, taking the time derivative of
both the sides of the previous equation \equ(302) at $t=0$ we obtain:
$$
 \big(T_{h_\e}\dot\Phi^\e_0\big)
 \  v^+_\e
 -
 {1\over \t_\e}{\det(g)\over 4}\big(\LL_3v^+_\e\big)
 =
  L_\e\cdot v^+_\e \Eq(303)
$$
We now write $v^+_\e$ as $v_\e^+=w^++\sum_{a=0,3,-}\d V_\e^a w^a$.
Projecting along the direction $w^+$, calling
$\FF^{,+}\defi \LL_+\FF$ and defining $\FF^{\a}$ such that 
$\FF=\sum_{\a=0,3,\pm}\FF^{\a}w ^\a$
and  $\FF^{\a,+}$ such that $\FF^{,+}=\sum_{\a=0,3,\pm}\FF^{\a,+}w ^\a$,
after some lengthly but straightforward algebra, reported in the Appendix,
we get 
$$
 \eqalign{
 &\d L_\e
 ={\det(g)\over 4}
 \Big[
 \e \FF^{+,+}(g)
 -\d\t_\e(g)
-\PP^+_\e\big(\d V_\e^0,\d V_\e^3,\d V_\e^-,\d L_\e)\Big]
}\Eq(303a)
$$
while, projecting along the other directions, we get
$$
 \eqalign{
 \big[\LL_3-(\l^a-\l^+)\big]\d V^a(g)
 =&
 \e \FF^{a,+} (g)
 -\d_{a,3}2\d V^0_\e(g)
 +\PP_\e^a\big(\d V_\e^0,\d V_\e^3,\d V_\e^-,\d L_\e)
}\Eq(303b)
$$
where $\{\PP^\a_\e\}_{\a=0,\pm,3}$ can be written as in \equ(DD). In
order to solve \equ(303a) and \equ(303b), as for \equ(202), we
first need to replace $2\d_{a,3}\d V^0_\e$ in the r.h.s.  of
\equ(303b), with the expression obtained by implicitly solving
the equation for $a=0$:
$$\d V^0_\e(g)=\int_{-\infty}^0\! ds\ 
e^{s\l^+}\lft[
 \e \FF^{0,+} +
\PP^0_\e\rgt]\circ\Phi_s\;,
$$
for $\PP^0_\e\circ\F_s\defi \PP_\e^a\lft(\{\d V_\e^a\circ \F_s\}_{a=0,3,-},\d L_\e\circ \F_s\rgt)$.
Substituting into \equ(303b), we get
$$
 \eqalign{
 \big[\LL_3-(\l^a-\l^+)\big]\d V^a
 =&
 \e \tilde\FF^{a,+}
 +\tilde\PP_\e^a\big(\d V_\e^0,\d V_\e^3,\d V_\e^-,\d L_\e)
}\Eq(304)
$$
for suitable $\{\tilde\FF^{a,+}\}_{a=0,}$ which 
depend  neither on $\{\d V^a_\e\}_{a=0,-.3}$, nor 
on $\d L_\e$; and it is linear in $\e$, but only 
H\"older continuous in $g$. While 
$\{\tilde\PP_\e^a\}_{a=0,-,3}$
are analytic in their arguments, 
and can be written as in 
\equ(DD), for suitable constants 
$\{\tilde C_{j,a}\}^{j=1,\ldots,4\atop a=0,-,3}$.
\*
\*

\sub(ECC){\it Existence of the perturbed unstable direction.}
Calling $f^0=\d V^0_\e$, $f^3=\d V^3_\e$, $f^-=\d V^-_\e$
and   $f^+=\d L^+_\e$, we can look for a solution of the 
\equ(303a) and \equ(303b) in the Banach space $\BB$ 
introduced in section 3.4. 
Again we introduce the operator:
$$
\big(M f\big)^\a
\defi
\cases{f^\a & if $\a=+$\cr
 \LL_3-(\l^a-\l^+)& if $\a=-,0,3$ ,}$$
and the function
$$T^\a_\e \big(f\big)
\defi\cases{
\e {\det\over 4}\FF^{+,+}
 -\d\t_\e{\det(g)\over 4}+\PP^+_\e(f^0,f^3,f^-,f^+)
& for $\a=+$ \cr\cr
 \e \tilde\FF^{a,+} 
 +\tilde\PP_\e^a(f^0,f^3,f^-,f^+)
 &for $\a=-,0,3$ ,}\Eq(Qbis)
$$
and we prove the following lemma.
\*
\*

{\bf\0Lemma 2.}
{\it There exists $\bar \e>0$ such that, for any $|\e|\le \bar\e$,
the equation 
$$
 Mf=T_\e (f)\Eq(th2)
$$
has unique solution in the ball of $\BB$ 
of radius $\e C$, for a suitable $C$.
Such a solution 
 is analytic in
$\e$.
}
\*

{\bf\0Proof.} It follows from arguments similar to those used in the
proof of Lemma 1. 
\hfill\qed\hskip1em\null
\*
Clearly the perturbed stable direction and Ljapunov exponent 
can be constructed in the very same way.
\*
\*
\section(SM, Proof of Theorem 3.)
   
\sub(PMSD){\it Markov Partition}.
It is worthwhile to remark that for topologically mixing 
Anosov flows the foliations $E^+$ and $E^-$ are not jointly 
integrable and therefore it is not possible to find a surface which 
contains a finite piece of the stable and unstable manifold 
of a given point (see \cita{Pl}).
That is why the following construction of
the Markov partition, \cita{B1} and \cita{R}, is slightly different 
from a naive generalization 
of the Markov partitions for the diffeomorphisms.
Fixed $\d>0$,
we define the local weak-stable and weak-unstable manifolds
passing through $g$ as
$$
W^{3,\pm}_\d (g) \defi \lft\{\big(\F_t\circ\F^\pm_\zeta\big)(g):|\zeta|,|t|<\d\rgt\}\;;
$$
both manifold are clearly $C^\o$.
Let $D$ be any closed $C^\o$ disk of dimension 2, transverse in each 
point  to the flow $F$.
Given two close points on $D$, $g,g':d(g,g')\le \a_1$, 
if $\a_1$ is small enough,
$$
\la g,g'\ra_D\defi
 W^{3,-}_\d(g)\cap W^{3,+}_\d(g') \cap D\Eq(AH)
$$
consists of one point.  We will say that $T$ is a rectangle on $D$
if $\lg g,g'\rg_D\in T$ for any $g,g'\in T$. 

The manifolds
$W^-_T(g)=\lft\{ \lg g,g'\rg_D : g'\in T\rgt\}$
and $W^+_T(g)=\lft\{ \lg g',g\rg_D : g'\in T\rgt\}$
are the projection of the stable and of the unstable manifolds
through $g$ on the rectangle $T$, which  can be seen as:
$$
T\=\la W^+_T(g),W^-_T(g)\ra\;. \Eq(AL1)
$$
Given a family of closed rectangle $\{T_1,\ldots,T_N\}$ on disks
$\{D_1,\ldots,D_N\}$ such that $T_i\subset {\rm int} D_i$ and
$T_i=\overline{{\rm int} T_i}$, we will call it a {\it proper family of
rectangles} if there exists $\a>0$ such that 
$\GG_\EE=\bigcup_{j=1}^N\bigcup_{t\in[0,\a]}\F_{-t}\lft(T_j\rgt)$; 
and for any $i\neq j$ at least one of the sets
$D_i \bigcap \left[\bigcup_{t\in[0,\a]}\Phi_t(D_j)\right]$ and
$D_j\bigcap\left[\bigcup_{t\in[0,\a]}\Phi_t(D_j)\right]$ is empty. 

We consider the points in $\bigcup_{j=1}^N T_j$ and 
define the {\it ceiling function}, $\th:\bigcup_{j=1}^N T_j\to\RRR_+$, as
the smallest positive time required for $\Phi_t(g)$ to cross
$\bigcup_{j=1}^N T_j$; and the {\it Poincar\'e map}, 
$\HH: \bigcup_{j=1}^N T_j \to  \bigcup_{j=1}^N T_j$, 
given by  $\HH(g)= \Phi_{\th(g)}(g)$.

Finally, 
the proper family of rectangles, $\{T_1,\ldots,T_N\}$, is called 
{\it Markov partition} if it satisfies two conditions:
for any $g\in T_i$, $\HH(g)\in T_j$
one has $\HH(g')\in T_j$ for any $g'\in W^-_{T_i}(g)$; and 
for any $g\in T_i$, $\HH^{-1}(g)\in T_j$
one has $\HH^{-1}(g')\in T_j$ for any $g'\in W^+_{T_i}(g)$.
In particular, the above construction 
of the rectangles $\{T_1,\ldots,T_N\}$ 
gives a Markov partition  for our dynamics.
\\

\sub(sd){\it Symbolic dynamics}

A characterization of the  SRB measure for the Anosov flows can be given in terms of 
symbolic dynamics.

Let $A$ be the incidence matrix associated with $\HH$,
i.e. 
$$
A_{i,j}=\cases{1& if ${\rm int}T_i \bigcap \HH({\rm int}T_j)\neq\emptyset$\cr
               0& otherwise.}
$$
Since the dynamics is mixing, there exists an integer $k$ such that
the matrix $A^k$ has only non-zero entries.
Accordingly,  we introduce the space of sequences 
$$
\Si_A\defi\lft\{\us\in \{1,\ldots,N\}^{\zzz}:A_{\s_i,\s_{i+1}}=1, i\in\ZZZ\rgt\}\;,
$$
the {\it shift map}, $\r:\Si_A\to\Si_A$, such that $(\r\us)_j=\s_{j+1}$
and the {\it coding map}, $X:\Si_A\to\bigcup_{j=1}^N T_j$, 
such that $X(\us)\defi \bigcap_{i=-\io}^{+\io} \overline{\HH^{-i}({\rm int}T_{\s_i})}$.
We remark that $\HH\circ X=X\circ \r$; furthermore,
endowing the space $\Si_A$ with the distance 
$|\us-\us'|=e^{-\n(\us,\us')}$,
for $\n(\us,\us')\defi\max\lft\{n\in\NNN:\s_i={\s'}_i\ \forall i:|i|\le n\rgt\}$,
the map $\r$ is continuous, and $X$ is H\"older continuous.

Finally, the coding is inherited by all $g\in \GG_\EE$:
after calling
$$Y\defi\big\{(\us,t)\in \Si_A\times\RRR_+: 0\le t\le(\th\circ X)(\us)\big\}$$
and identifying $(\us,(\th\circ X)(\us))$ with $(\r\us,0)$,
let $q:Y\to \GG_\EE$
be the one-to-one map defined by  $q(\us,t)=\big(\F_{t}\circ X\big) (\us)$; 
then
$$
\big(\F_t \circ q\big)(\us,s)=q\big(\r^k\us,t'\big)\Eq(sf)
$$
for the unique $k$ such that $t'\defi t+s-\sum_{j=0}^{k-1}\big(\th\circ X\circ \r^j\big)(\us)$
satisfies $0\le t'<\big(\th\circ X\circ \r^k\big)(g)$.
\\

\sub(SRB){\it SRB measure}.
Given a  H\"older continuous $f:\Si_a\to \RRR$, there is a standard procedure
to construct the {\it equilibrium state}, 
an $\r$-invariant, Gibbs measure on $\Si_A$, $\n_f$: we do not 
give here the details, see \cita{B2} for proof and details; we 
only state that such measure is the unique Gibbs measure 
with formal Hamiltonian 
$$
 H(\us)\defi \sum_{j=-\io}^{+\io}f(\r^j\us)\;.\Eq(fh)
$$
Now, let $\L^+_t(g)$ be the Jacobian of the linear map 
$T\F_t:E^+_g\to  E^+_{\F_t(g)}$;
and let 
$$
\l^+(g)\defi -\lft.{d\ln \L^+_t(g)\over d t}\rgt|_{t=0}
$$
which exists and is analytic in $g$.
As proved in \cita{BR} (theorem 5.1), the SRB measure  defined  
in \equ(srb) is given by
$\m(\OO)=\n_{\hat f^+\circ X}(\widehat\OO\circ X)$, for
$\hat f^+,\widehat\OO:\bigcup_{j=1}^N T_j\to \RRR$ such that 
$$
\hat f^+(g)\defi \int_0^{\th(g)}\!ds\ 
\big(\l^+\circ\F_s\big)(g)\;,\qquad
\widehat\OO(g)\defi \int_0^{\th(g)}\!ds\ 
\big(\OO\circ\F_s\big)(g)\;.
$$
Since $\F$ is a Hamiltonian flow, $\m$ is the 
Lebesgue measure. 

For the perturbed, non-Hamiltonian flow, $\F_t^\e$, the SRB measure
is generally not absolutely continuous w.r.t. the Lebesgue measure.
It is possible to 
use the conjugation $h_\e$ to construct such a measure;
we remark, anyway, that contrary to the naive expectation,
the rectangles $\{\tilde T_1^\e,\ldots,\tilde T_N^\e\}$
such that $\tilde T_j^\e\defi h_\e(T_j)$ do not yield
a Markov Partitions, since the rectangles are no longer portions
of differentiable manifolds.

In order to simplify the forthcoming construction, we observe that 
the each of the discs $\{D_1,\ldots,D_N\}$ can be thought as a closed
piece of the 2-dimensional, $C^\o$ manifold obtained by intersecting
a certain $C^\o$ surface, $\SS\subset\GG$, transverse 
to the flow $\F$, and the $C^\o$ surface $\GG_\EE$.  

It is than natural 
to define the perturbed partition $\{T^\e_1,\ldots,T^\e_N\}$ 
such that $T^\e_j$ is the intersection of 
 $\bigcup_{t:|t|<\d}\F^\e_t(\tilde T^\e_j)$ with $\SS$: this gives
a $C^\o$ rectangle, since the boundaries of $T^\e_j$ are portions of 
the weak-stable and weak-unstable perturbed manifold, which are analytic since
the perturbed flow is Anosov.

In other words, if $\SS=\{g\in \GG:S(g)=0\}$ for a certain 
function $S:\GG\to\RRR$, let $s_\e:\GG_\EE^\e\to \RRR_+$
be the time such that 
$$
\big(S\circ\F^\e_{s_\e(g)}\big)(g)=0
$$
We will use as conjugation between the umperturbed and perturbed dynamics
the function $p_\e:\GG_\EE\to \GG_\EE^\e$
$$
p_\e(g)\defi \big(\F^\e_{(s_\e\circ h_\e)(g)}\circ h_\e\big)(g)
$$
which is clearly analytic in $\e$ and H\"older continuous 
in $g$.
Accordingly, the coding for the perturbed flow,
$X_\e:\Si_A\to\bigcup_{j=1}^N T^\e_j$ is given by 
$X_\e=p_\e\circ X$; and, if $\th_\e$ is
the perturbed return time, namely it is the shortest,
nonzero time
allowing to 
define the perturbed Poincar\'e map, 
$\HH_\e:\bigcup_{j=1}^N T^\e_j\to\bigcup_{j=1}^N T^\e_j$,
as 
$\HH_\e(g)\defi\F^\e_{\th_\e(g)}(g)$,
the SRB measure average of
the H\"older continuous function
$\OO$ is given by
$$\m_\e(\OO)=\n_{\hat f^+_\e\circ X_\e}(\widehat\OO_\e\circ X_\e)\Eq(srbe)$$ 
for $\hat f^+_\e,\widehat\OO_\e:\bigcup_{j=1}^N T_j\to \RRR$ such that 
$$\eqalign{
 \hat f_\e^+(g)\defi 
&\int_{0}^{\th_\e(g)}\!ds\ 
\big(\l_\e^+\circ\F^\e_s\big)(g)\;,\qquad
\widehat\OO_\e(g)\defi \int_0^{\th_\e(g)}\!ds\ 
\big(\OO\circ\F^\e_s\big)(g)\;.
}$$
We want to prove that, the conjugation $p_\e$, thought essential 
to construct the SRB, plays no role in the actual computation of
the mean values: conjugating the dynamics with $h_\e$ would have
given (formally) the same result.
Indeed, using the identity $\big(\F^\e_s\circ p_\e\big)(g)=\big(\F^\e_{s+(s_\e\circ h_\e)(g)}
\circ h_\e\big)(g)$;
and, for $\tilde \th_\e:\bigcup_{j=1}^N \tilde T^\e_j \to \RRR_+$
the first return time to  the H\"older continuous 
manifold $\bigcup_{j=1}^N \tilde T^\e_j$,
the identity 
$\big(\th_\e\circ p_\e\circ h_\e\big)(g)=
\big(\tilde\th_\e\circ h_\e\big)(g) + 
\big(s_\e\circ h_\e\circ \HH\big)(g)- \big(s_\e\circ h_\e\big)(g)$,
holds true. And therefore
$$\eqalign{
\big(\hat f^+_\e\circ p_\e\big)(g)
&=\int_0^{(\th_\e\circ p_\e)(g)}\!ds\ 
\big(\l_\e^+\circ \F^\e_s\circ p_\e\big)(g)
\cr&
=
\int_{(s_\e\circ h_\e)(g)}^{(\tilde\th_\e\circ h_\e)(g)+ (s_\e\circ h_\e\circ \HH)(g)}
\!ds\ 
\big(\l_\e^+\circ \F^\e_s\circ h_\e\big)(g)
\cr&\defi
\int_{0}^{(\tilde\th_\e\circ h_\e)(g)}
\!ds\ 
\big(\l_\e^+\circ \F^\e_s\circ h_\e\big)(g)+\big(\widehat F^+_\e\circ \HH\big)(g)-\widehat F^+_\e(g)
}\Eq(coc)$$
for a suitable, H\"older continuous function $\widehat F^+_\e:\bigcup_{j=1}^N T_j \to\RRR$.
Now, it is well known that
the {\it cocicle} $\big(\widehat F^+_\e\circ \HH\big)(g)-\widehat F^+_\e(g)$ does not 
count in the computation of  the equilibrium states: since a very similar 
development can be done  for $\big(\widehat \OO_\e\circ p_\e\big)(g)$,
we can make the following replacement in the expression 
for the SRB measure in \equ(srbe):
$$\eqalign{
&\big(\hat f^+_\e\circ p_\e\big)(g)\longrightarrow 
\int_{0}^{(\tilde\th_\e\circ h_\e)(g)}
\!ds\ 
\big(\l_\e^+\circ \F^\e_s\circ h_\e\big)(g)\;,
\cr&
\big(\widehat \OO_\e\circ p_\e\big)(g)
\longrightarrow 
\int_{0}^{(\tilde\th_\e\circ h_\e)(g)}
\!ds\ 
\big(\OO\circ \F^\e_s\circ h_\e\big)(g)\
}$$
We conclude the proof of Theorem 3 noticing 
that both above expressions are analytic in $\e$: this is 
because of the identities $\big(\tilde\th_\e\circ h_\e\big)(g)=T^\e_{\th(g)}(g)$ and
$\big(\F^\e_{T^\e_s(g)}\circ h_\e\big)(g)=\big(h_\e\circ \F_s\big)(g)$;
by the analyticity of $\l_\e^+\circ h_\e=L^+_\e$ and $T^\e_s$; and finally because,
thanks to the  hypothesis that $\OO$ is analytic, $\OO\circ h_\e$ is analytic too. 
\*
\*
\section(CO, Conclusion and outlook.)

The geodesic motion of surfaces with constant negative curvature 
has been are the typical example of Anosov system: in \cita{A}
their {\it structural stability}, namely the existence of the 
conjugation between two close flows, was first proved. 
Later on, in  \cita{KKPW} and in \cita{LMM} (in particular in appendix A)
very general results on the regularity of 
$h_\e$ in $\e$ where proved using the the contracting mapping theorem
or implicit function theorem,
a point of view introduced by Moser, \cita{Mo} and Mather\cita{Ma}. 
Anyway, in all above papers only the case 
$\GG^\e_\EE=\GG_\EE$ has been considered.

Our technique is more in the 
spirit of \cita{BKL} (see also \cita{BFG} and \cita{GBG}). 
While \cita{KKPW} discusses the regularity  
of the {\it topological entropy} of the system, 
and consequently of the ``equilibrium states'' associated to a generic H\"older continuous
``potential'', the final address of the present paper has been to  study 
the analyticity of a special equilibrium state, the SRB measure. In 
order to do it, we constructed and proved analyticity in $\e$ 
of the contraction rate of the unstable phase space.
\*
\*

\appendix(A1, Explicit computations.)

\asub(a1){\it Explanation of \equ(202)}.
Taking the time derivative in $t=0$, the l.h.s. member
of the first equation in \equ(conj) gives
$$
{\det(g)\over 4}\lft[w^3(g) + \sum_{\a=0,\pm,3} \d h_\e^\a(g) (\LL_3 w^\a)(g)+
\sum_{\a=0,\pm,3} (\LL_3\d h_\e^\a)(g) w^\a(g)\rgt] 
$$
Therefore \equ(202) follows from the identity 
$\big(w^3\circ h_\e\big)(g)=w^3+ 
\sum_{\a=0,\pm,3}\d h^\a_\e(g) (\LL_\a w^3)(g)$ and 
from \equ(anz1), which gives
$$ 
 {\big(\det\circ h_\e\big)(g)\over\det(g)}
= 1 -2\d h_\e^0(g) +(\d h_\e^0)^2(g) -(\d h_\e^3)^2(g)- \d h_\e^+(g)\d h_\e^-(g).
$$
from which \equ(202) for suitable functions $\{\RR^\a_\e\}_{\a=0,\pm,3}$.
\*
\*

\asub(2){\it Explanation of Explanation of \equ(303a) and \equ(303b)}.
Using th decomposition for $v_\e$ after \equ(303),
\equ(303) reads:
$$
 \eqalign{
 &\big(\LL_+\dot\F^\e_0\big)(g)
 +\sum_{a=0,3,-}\big(\LL_a\dot\Phi^\e_0\big)(g)\d V^a(g)
 -
 {1\over \t_\e(g)}{\det(g)\over 4}\big(\LL_3w^+\big)(g)\cr&
 -
 {1\over \t_\e(g)}{\det(g)\over 4}\sum_{a=0,3,-}\d V^a(g)\big(\LL_3w^a\big)(g)-
 {1\over \t_\e(g)}{\det(g)\over 4}\sum_{a=0,3,-}w^a(g)\big(\LL_3\d V^a\big)(g)\cr
&
 =
 L_\e(g) w^+(g) + L_\e(g)\sum_{a=0,3,-}\d V^a(g) w^a (g)+ 
\big(T_{h_e(g)}\dot\F_0^\e- T_{g}\dot\F_0^\e\big)v_\e (g)}\Eqa(303bisbis)
$$
By \equ(Hebis), and neglecting the dependence in $g$,
$$
 \eqalign{
 &{\det\over 4}\big(\LL_+w^3\big)
+{\det\over 4}\sum_{a=0,3,-}\big(\LL_aw^3\big)\d V^a
+{\det\over 2}w^3\d V^0\cr&
 -
 {1\over \t_\e}{\det(g)\over 4}\big(\LL_3w^+\big)
 -
 {1\over \t_\e}{\det(g)\over 4}\sum_{a=0,3,-}\d V^a\big(\LL_3w^a\big)-
 {1\over \t_\e}{\det(g)\over 4}\sum_{a=0,3,-}w^a\big(\LL_3\d V^a\big)\cr&
 =
 L_\e\cdot  w^+ + L_\e\sum_{a=0,3,-}\d V^a w^a\cr &
 -\e {\det\over 4}\big(\LL_+\FF\big)
 -\e{\det\over 4}\sum_{a=0,3,-}\big(\LL_a\FF\big)\d V^a
 -\e{\det\over 2}\FF\d V^0+ 
\big(T_{h_e(g)}\dot\F_0^\e- T_{g}\dot\F_0^\e\big)v_\e (g)
}\Eqa(303null)
$$
With further developments;
using the identity following \equ(Hbis)
and the decomposition $L_\e=\l^++\d L_\e$
$$
 \eqalign{
 &\d\t_\e\big(\LL_+w^3\big)
-\sum_{a=0,3,-}\big(\LL_3\d V^a-(\l^a-\l^+)\d V^a\big) w^a
+2w^3\d V^0\cr&
 =
 {4\over \det}\d L_\e\cdot  w^+ 
 -\e \big(\LL_+\FF\big)+\PP_\e\big(\d V_\e^0,\d V_\e^3,\d V_\e^-,\d L_\e)\;.
}\Eqa(303ter)
$$
Projecting along the direction $w^+$, calling
$\FF^{,\a}\defi \LL_\a\FF$ and defining $\FF^{\a},\PP^\a$ such that 
$\FF=\sum_{\a=0,3,\pm}\FF^{\a}w ^\a$ and similarly for $\PP^\a$;
finally defining   $\FF^{\a,\b}$ such that 
$\FF^{,\b}=\sum_{\a=0,3,\pm}\FF^{\a,\b}w ^\a$,
we get 
\equ(303a) and \equ(303b).
\vskip1cm
{\bf Acknowledgments}
P.F. gratefully acknowledges
the hospitality and the financial 
support of the Erwin Schr\"odinger 
Institute for Mathematical Physics (Vienna)
during the 
preparation of this work.

\*

\rife{A}{A}{Anosov D.:
 {\it Geodesic flows on closed Riemannian manifolds with negative curvature},
 { Proc.Steklov Inst.Math.} {\bf 90} (1967).}
\rife{B1}{B1}{Bowen R.:
 {\it Symbolic dynamics for hyperbolic flows},
 { Amer.J.Math.} {\bf 95} { 429-459} (1972).}
\rife{B2}{B2}{Bowen R.:
 {\it Equilibrium states and ergodic theory for Anosov diffeomperphism},
 { Lecture Notes Math Vol 470, Springer}  (1973).}
\rife{BR}{BR}{Bowen R., Ruelle D.:
 {\it The ergodic theory for Axiom A flows},
 { Invet.Math.} {\bf 29} { 181-202} (1975).}
\rife{BFG}{BFG}{Bonetto F., Falco P., Giuliani A.:
 {\it Analyticity of the SRB measure of a lattice of coupled Anosov diffeomorphisms of the torus},
 { J.Math.Phys.} {\bf 45} { 3282-3300} (2004)}
\rife{BGM}{BGM}{Bonetto F., Gentile G., Mastropietro V.:
 {\it Electric fields on a surface of constant negative curvature},
 { Ergod.Th.Dynam.Sys.} {\bf 13} { 681-696} (2000). }
\rife{BKL}{BKL}{Bonetto F.,  Kupiainen A. and Lebowitz J.:
 {\it Absolute continuity of projected SRB measures of coupled 
Arnold cat map lattices},
 { Ergod.Th.Dynam.Sys.} {\bf 25} { 59-88}(2005). }
\rife{CEG}{CEG}{Collet P., Epstein H. and  Gallavotti G.:
 {\it  Perturbation of geodesic flows on surfaces of constant
 negative curvature and their mixing properties},
 {\  Comm. Math.Phys.} {\bf 95} { 61--112} (1984). }
\rife{ECM}{ECM}{Evans D.J., Cohen E.G.D. Morriss G.P.:
 {\it Probability of second law violations in shearing steady flows},
 { Phys.Rev.Lett.} {\bf 71} { 2401-2404}(1993). }
\rife{F}{F}{Ford L.:
 {\it Automorphic functions},
 { Chelsea Publishing Company} (1951). }
\rife{GBG}{GBG}{Gallavotti G., Bonetto F., Gentile G.:
 {\it Aspects of ergodic, qualitative and statistical theory of motion},
 { Springer} (2004).}
\rife{GC}{GC}{Gallavotti G., Cohen E.G.D.:
 {\it Dynamical ensembles in non equlibrium statistical mechanics},
 { Phys.R.Lett.} {\bf 74} { 2694-2697} (1995).}
\rife{G}{G}{Gallavotti G.:
 {\it Reversible Anosov diffeomorphisms and large deviations},
 { Math. Phys.Elect.J.} {\bf 1} { 1-12} (1995).}
\rife{Ge}{Ge}{Gentile G.:
 {\it A large deviation theorem for Anosov flows},
 { Forum Math.} {\bf 10} { 89-118} (1998). }
\rife{GF}{GF}{Gelfand I., Fomin S.:
 {\it Geodesic flows on a manifolds of constant negative curvature},
 { Am.Math.Soc.Tr.} {\bf 1}, { 49} (1965). }
\rife{GGP}{GPP}{Gelfand I.M., Graev M.I., Piateckii-Shapiro I.I.:
 {\it Representation theory and automorphic functions},
 { Saunders}, (1969). }
\rife{GZG}{GZG}{Giuliani A., Zamponi F., Gallavotti G.:
 {\it Fluctuation Relation beyond Linear Response Theory},
 { J.Stat.Phys.} {\bf 119} { 909-944} (2005).}
\rife{KKPW}{KKPW}{Katok A., Knieper G., Pollicot M., Weiss H. :
 {\it Differentiability and analyticity of topological entropy for Anosov and geodesic flow},
 { Invent.Math.} {\bf 1}  (1882).}
\rife{LMM}{LMM}{de la LLave R., Marco J.M. Moryon R.:
 {\it Canonical perturbation theory
 of Anosov systems and regularity results for the Livsic cohomology equation},
 { Ann. Math.} {\bf 123} { 537-611} (1986).}
\rife{Ma}{Ma}{Mather J.: appendix to \cita{Sm}}
\rife{Mo}{Mo}{Moser J.:
 {\it On a theorem of Anosov},
 { J.Differ.Equat.} {\bf 5} { 411-440} (1969).}
\rife{Pl}{Pl}{Plante J.F.:
 {\it Anosov flows},
 { Amer.J.Math.} {\bf 94} { 729-754} (1972).}
\rife{P}{P}{Poincar\'e H.:
 {\it Automorphic functions},
 { Acta Math.} {\bf 1}  { 1-} (1882). }
\rife{R}{R}{Ratner R.:
 {\it Markov partition for Anosov flows onn-dimensional manifolds},
 { Isr.J.Math.} {\bf 15} { 92-114} (1973).}
\rife{S}{S}{Sinai Ya.G.:
 {\it Gibbs measure in ergodic theory},
 { Russ.Math.Surveys} {\bf 166}  { 21-69}(1972).}
\rife{Sm}{Sm}{Smale S.:
 {\it Differentiable dynamical systems}
 { Bull.Amer.Math.Soc.} {\bf 73} { 747-817} (1967)}

\biblio

\end